# Performance Analysis of Small Cells' Deployment under Imperfect Traffic Hotspot Localization

Aymen Jaziri *+, Ridha Nasri *, Tijani Chahed +
*Orange Labs, +Telecom Sudparis
Email: {aymen.jaziri, ridha.nasri}@orange.com, tijani.chahed@telecom-sudparis.eu

*Abstract*—Heterogeneous Networks (HetNets), long been considered in operators' roadmaps for macrocells' network improvements, still continue to attract interest for 5G network deployments. Understanding the efficiency of small cell deployment in the presence of traffic hotspots can further draw operators' attention to this feature. In this context, we evaluate the impact of imperfect small cell positioning on the network performances. We show that the latter is mainly impacted by the position of the hotspot within the cell: in case the hotspot is near the macrocell, even a perfect positioning of the small cell will not yield improved performance due to the interference coming from the macrocell. In the case where the hotspot is located far enough from the macrocell, even a large error in small cell positioning would still be beneficial in offloading traffic from the congested macrocell.

## I. INTRODUCTION

Heterogeneous Networks (HetNets), composed of small cells (micro, pico, Remote Radio Heads (RRH), relay and femto cells) within macro-cell coverage, have been shown to be efficient in improving network performance [1], [2], mostly by covering traffic HotSpots (HS) and coverage holes. The resulting enhanced network performance depends among other parameters on the location of the small cells deployment, and this is the focus of our present work where we study the impact of imperfect small cells deployment on the overall network performance.

The analysis of Small Cell (SC) deployment was the subject matter of many works but mostly using simplified network topology models which yields either optimistic or pessimistic results. For instance, authors in [1]–[3] considered a HetNet network with different tiers to evaluate different performance metrics such as the coverage probability, the average achievable rate and the average load per tier. The network structure in each tier is based on a spatial Poisson Point Process (PPP) which is well suited for SC networks, where the Base Station (BS) positions are more irregular than for the Macro Cell (MC) layout. Each tier differs from the other by the average transmit power, the base station density and the supported data rate. Using PPP model allows for simple and tractable analytical expressions but it remains quite different from the real network layout, mainly for macro tier which is close to the hexagonal grid with imperfections [4]. Moreover, the deployment of SCs should be made based on the traffic distribution as well. Considering small cell distribution as a PPP model implies that the traffic HS - in case it is the reason for deploying SCs - follows also a PPP model with a constant traffic value in each HS which is not the case, as shown in [5]. This random spatial model can however be improved by incorporating new parameters such as the minimum separation distance between BSs and/or the spatial traffic distribution.

A different approach was used in [6] where the authors used a fluid model in order to study the impact of SC location on the performance and Quality of Service (QoS) in HetNets. A closed-form expression of the Signal plus Interference to Noise Ratio (SINR) is found for each position in the network and is used, along with the throughput distribution in both MC and SC layers, in order to assess the impact of SC location on the network performance. The quite regularity of real networks and non-homogeneous spatial traffic distribution are however not taken into consideration in reference [6].

In general, it is recommended to perform analysis in a hexagonal network layout because of the fact that radio engineers start the design of the network based on a hexagonal model. The deviation of the real network structure from the hexagonal model is basically linked to the constraints imposed by engineering rules, field imperfections and government charters. Following this approach, we derive and evaluate, in the present work, several performance metrics in an infinite hexagonal MC network for three different scenarios: with MCs only, with perfectly deployed SCs and with SCs deployed with considering imperfections in the positioning. The former does not involve any SC deployment and so interference comes only from neighboring MCs. The second scenario assumes that a SC is perfectly deployed in the peak of the traffic HS. The last scenario involves errors in the positioning of SCs relative to the position of the traffic HS.

The main contributions of this paper are twofold. First, we evaluate the gain that can be generated from the deployment of a SC considering an error in HS localization, as compared to perfect HS localization. Second, we identify the threshold of HS localization errors that can be tolerated in operational tasks mainly for HetNet design.

The remainder of this paper is organized as follows. In the next section, we present the downlink system model. In section III, we detail the performance analysis for the above-mentioned scenarios. Numerical results are highlighted in section IV, and, in section V, we conclude with a brief discussion of the results of this paper.

## II. DOWNLINK SYSTEM MODEL

### A. System setup and BS location model

We consider a cellular network with an infinite number of omni-sectorial MCs, each one transmitting with power level $P$. The location of the MCs is drawn following a hexagonal grid layout (see Fig. 1) with inter-site distance denoted by $\delta$. Next, we add a SC located in $(R_s, \theta_s)$ as illustrated in Fig. 1. The transmit power level of the SC is $P_s = \alpha P$ with $\alpha < 1$.

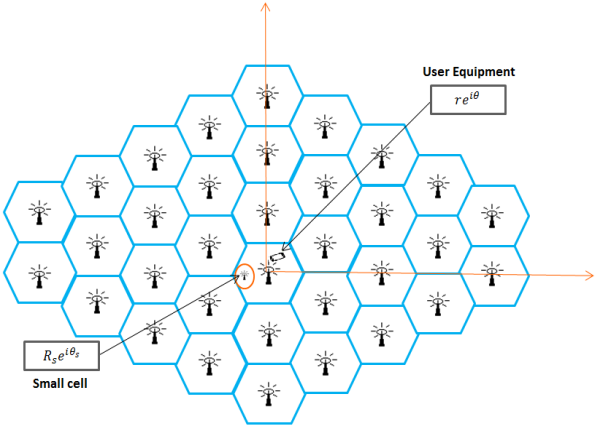

Fig. 1: Network layout.

A given User Equipment (UE) with polar coordinates $m = (r, \theta)$ is served by either the central MC or the deployed SC, depending on the relative signal strength coming from both antennas, and the rest of the cells play the role of interfering ones.

In order to evaluate the efficiency of deploying SCs, we consider a traffic HS with polar coordinates $(R_h, \theta_h)$. Without loss of generality, we assume that this HS is located inside the central MC of the network. This means that $R_h$ is smaller than the radius of the cell $R = \delta\sqrt{\frac{\sqrt{3}}{2\pi}}$ defined by the radius of the disk having the same area as the hexagon[1]. Then, a Gaussian distribution defines the UE location distribution inside the HS and its measure is given by

$$dt(r,\theta) = \frac{1}{2\pi\sigma^2} e^{-\frac{r^2+R_h^2-2rR_h\cos(\theta-\theta_h)}{2\sigma^2}} r\,dr\,d\theta \quad (1)$$

$\sigma$ represents the standard deviation of the distribution. In simulations, $\sigma$ is equal to $0.2$ and an example of the relative spatial distribution is plotted in Fig. 2.

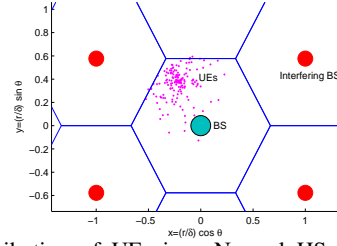

Fig. 2: Distribution of UEs in a Normal HS with $\sigma = 0.2$ in $(R_h = 0.44\delta, \theta_h = \frac{2\pi}{3})$.

As indicated in the introduction, we consider three scenarios: In Scenario 1, the network is composed of MCs only. This scenario represents a benchmark allowing the comparison of a network containing SCs with a network without SCs. Scenario 2 adds one SC inside the central MC. This SC is perfectly deployed and its position matches exactly with the HS's position. In Scenario 3, we consider a SC deployed inside the central MC as in Scenario 2 but we introduce some errors in the HS localization so that the positions of the HS and the SC are not the same.

To model the wireless channel, we consider a distance based pathloss metric with a standard function given by $a|m - C|^{-2b}$, where $|m - C|$ is the distance between the UE $m$ and any cell $C$ in the network, which can be either a MC or a SC. $a$ is a pathloss constant[2] which depends on the type of the environment relative to the type of the cell (indoor, outdoor, rural, urban...) and $2b > 2$ is the pathloss exponent coefficient.

Based on the proposed pathloss model, the UE located in $m = (r, \theta)$ is served by the SC only if the RSRP (Reference Signal Received Power) [9] of the SC is higher than the RSRP coming from the MC. This can be expressed by the following inequality

$$P_s|re^{i\theta} - R_s e^{i\theta_s}|^{-2b} > Pr^{-2b} \quad (2)$$

If the constraint in (2) is not satisfied, the UE will be connected to the MC.

Based on inequality (2), we evaluate the absorption coefficient (denoted by $\mu$) which reflects the percentage of mobile locations (generated according to a given traffic distribution) that can be served by the SC. This performance metric is given by

$$\mu = \frac{1}{S_0} \int_S \mathbb{1}\left(P_s|re^{i\theta} - R_s e^{i\theta_s}|^{-2b} > Pr^{-2b}\right) dt(r,\theta) \quad (3)$$

where $S$ is the covered area by the central MC and the deployed SC.

$$S_0 = \int_S dt(r,\theta) \quad (4)$$

---

[1]The radio design based on the disk having the same area as the hexagonal cell, avoids the over-overlapping between cells and the appearance of coverage holes when the network is deployed.

[2]Without loss of generality, we consider that the transmit power levels $P$ and $P_s$ include as well the pathloss constant $a$, antenna gain, cable loss, UE antenna gain and body loss.

## B. SINR VS Throughput: Link level capacity curve

The SINR received at the UE and its throughput are respectively denoted by $\gamma(r,\theta)$ and $\eta(r,\theta)$ indexed with $m$ if it is received from the MC and with $s$ if it is received from the SC.

The relation between $\gamma$ (in linear scale) and $\eta$ (in Mbps) depends on the UE capacity, the available bandwidth, the radio conditions, the type of the service etc...
This relation is often modeled by a modified Shannon formula as stated in [10],

$$\eta = min(K_1 \times W \times \ln(1 + K_2 \times \gamma), \eta_0) \quad (5)$$

where $K_1$ and $K_2$ are two variables depending on transmission conditions foregoing and can be adapted for each UE; $W$ is the used bandwidth; $\eta_0$ is the maximum throughput representing the capability of the target UE category.
For the present paper, we consider UE category equal to 3 working at 20 MHz and we find out that, under labs measurement conditions, $\eta_0 = 98 Mbps$, $K_1 = 0.85$ and $K_2 = 1.9$.

## III. PERFORMANCE ANALYSIS

In the presence of a HS in the central MC, we define the mean user throughput (MUeTh) in the region $S$, covered by the MC and the deployed SC as follows

$$\eta = \eta_m + \eta_s \quad (6)$$

where

$$\eta_m = \frac{1}{S_0}\int_S \mathbb{1}\left(P_s|re^{i\theta} - R_s e^{i\theta_s}|^{-2b} < Pr^{-2b}\right) \times$$
$$min(K_1 W \ln(1 + K_2 \times \gamma_m(r,\theta)), \eta_0)dt(r,\theta) \quad (7)$$

$$\eta_s = \frac{1}{S_0}\int_S \mathbb{1}\left(P_s|re^{i\theta} - R_s e^{i\theta_s}|^{-2b} > Pr^{-2b}\right) \times$$
$$min(K_1 W \ln(1 + K_2 \times \gamma_s(r,\theta)), \eta_0)dt(r,\theta) \quad (8)$$

and $dt(r,\theta)$ is the measure representing the spatial traffic distribution reflecting the presence of a HS.
As mentioned in II-B, $\gamma_m$ and $\gamma_s$ are the SINRs received at a UE with polar coordinates $(r,\theta)$ and served respectively by the MC and the SC. Hence, $\gamma_m$ and $\gamma_s$ are expressed as follows

$$\gamma_m(r,\theta) = \frac{1}{f(r) + \alpha|re^{i\theta} - R_s e^{i\theta_s}|^{-2b}r^{2b} + \frac{P_N}{P}r^{2b}} \quad (9)$$

$$\gamma_s(r,\theta) = \frac{P_s|re^{i\theta} - R_s e^{i\theta_s}|^{-2b}}{(f(r)+1)Pr^{-2b} + P_N} \quad (10)$$

where $P_N$ is the thermal noise power and $f(r)$ represents the interference factor in a network composed of only MCs. It is defined by the ratio between the power coming from all the interfering MCs and the received power from the serving MC.
In order to evaluate the impact of infinite number of interfering MCs, we have established and validated in [7] an efficient and simple expression of the interference factor for the considered hexagonal network model. This formula is expressed as follows:

$$f(r) \approx 6x^{2b}\left(\frac{1+(1-b)^2 x^2}{(1-x^2)^{2b-1}} + \omega(b) - 1\right) \quad (11)$$

with

$$\omega(b) = 3^{-b}\zeta(b)\left(\zeta(b,\frac{1}{3}) - \zeta(b,\frac{2}{3})\right) \quad (12)$$

where $x = \frac{r}{\delta}$, $\zeta(.)$ and $\zeta(.,.)$ are respectively the Riemann Zeta and Hurwitz Riemann Zeta functions.

We notice that the performance analysis in Scenario 1 is equivalent to the case of Scenario 3 but with deploying a SC far enough from the MC and HS positions. For instance, $R_s$ can be taken as equal to $+\infty$. Hence, the inequality in (2) will be never satisfied for UEs located inside the HS and it follows that

$$\eta = \eta_m \quad and \quad \eta_s = 0 \quad (13)$$

On the other hand, Scenario 2 is illustrated by the equality

$$(R_s, \theta_s) = (R_h, \theta_h) \quad knowing \ that \quad R_h \leq R \quad (14)$$

Scenario 3 is eventually depicted by the following conditions

$$(R_s, \theta_s) \neq (R_h, \theta_h) \quad and \quad R_s \leq R \quad (15)$$

### A. Scenario 1: with macro-cells only deployment

As mentioned before, we take $R_s = \infty$ and so all the UEs in the HS, following $dt(r,\theta)$ defined in (1), are served by the MC. In this case, the mean user throughput is equal to

$$\eta = \frac{2}{S_0}\frac{K_1 W}{2}\frac{1}{\sigma^2}e^{-\frac{R_h^2}{2\sigma^2}} \times$$
$$\int_0^R re^{-\frac{r^2}{2\sigma^2}}\ln\left(1 + \frac{K_2}{max(\rho, g(r))}\right)\frac{1}{\pi}\int_0^\pi e^{\frac{rR_h cos(\theta)}{\sigma^2}} dr d\theta$$
$$= \frac{2}{S_0}\frac{K_1 W}{2}\frac{1}{\sigma^2}e^{-\frac{R_h^2}{2\sigma^2}} \times$$
$$\int_0^R re^{-\frac{r^2}{2\sigma^2}}\ln\left(1 + \frac{K_2}{max(\rho, g(r))}\right) I_0\left(\frac{rR_h}{\sigma^2}\right) dr \quad (16)$$

where $\rho$ is equal to $K_2\left(e^{\frac{\eta_0}{K_1 W}} - 1\right)^{-1}$, $I_0(.)$ is the first order of the modified Bessel function of the first kind [11] and

$$g(r) = f(r) + \frac{P_N}{P}r^{2b} \quad (17)$$

Function $g$ realizes a continuous increasing function from $[0, R]$ to $[0, g(R)]$. And so, it is possible to explicitly inverse $g$ using series reversion or also by switching numerically the axes $x$ and $y$. A simple and accurate inverse function of $g$ is provided in [8]. Using the monotonicity of $g$, the expression of $\eta$ is further simplified:

$$\eta = \frac{2}{S_0}\frac{K_1 W}{2}\frac{1}{\sigma^2}e^{-\frac{R_h^2}{2\sigma^2}}\ln\left(1 + \frac{K_2}{\rho}\right) \times$$
$$\int_0^{min(R, g^{-1}(\rho))} re^{-\frac{r^2}{2\sigma^2}} I_0\left(\frac{rR_h}{\sigma^2}\right) dr +$$
$$\frac{2}{S_0}\frac{K_1 W}{2}\frac{1}{\sigma^2}e^{-\frac{R_h^2}{2\sigma^2}} \times$$

$$\int\limits_{min(R,g^{-1}(\rho))}^{R} re^{-\frac{r^2}{2\sigma^2}} \ln\left(1 + \frac{K_2}{g(r)}\right) I_0\left(\frac{rR_h}{\sigma^2}\right) dr \quad (18)$$

### B. Scenario 2: with perfectly deployed small cell

In Scenario 2, UEs are served either by the MC or by the SC depending on the highest received RSRP. So, the mean throughput of UEs served by the MC is

$$\eta_m = \frac{2}{S_0}\frac{K_1 W}{2\pi}\frac{1}{\sigma^2}e^{-\frac{R_h^2}{2\sigma^2}}\int_0^R re^{-\frac{r^2}{2\sigma^2}}\int_{cos^{-1}(h(r)|_{-1}^1)}^{\pi} e^{\frac{rR_h cos(\theta)}{\sigma^2}}$$
$$\ln\left(1 + \frac{K_2}{max\left(\rho, g(r) + \alpha r^{2b}|re^{i\theta} - R_h|^{-2b}\right)}\right) drd\theta \quad (19)$$

where

$$h(r) = \frac{R_s^2 + r^2 - \alpha^{\frac{1}{b}}r^2}{2rR_s} \quad (20)$$

$$h(r)|_{-1}^1 = \begin{cases} h(r) \text{ if } -1 < h(r) < 1 \\ -1 \text{ if } h(r) < -1 \\ 1 \text{ if } h(r) > 1 \end{cases} \quad (21)$$

and $P_s|re^{i\theta} - R_s e^{i\theta_s}|^{-2b} < Pr^{-2b}$ is only verified when $cos(\theta - \theta_s) < h(r)$.

$$cos^{-1}(h(r)|_{-1}^1) = \begin{cases} cos^{-1}(h(r)) & \text{if } r_1 < r < r_2 \\ 0 & \text{otherwise} \end{cases} \quad (22)$$

with $r_1 = \frac{R_s}{1+\alpha^{\frac{1}{2b}}}$ and $r_2 = \frac{R_s}{1-\alpha^{\frac{1}{2b}}}$.

And so we obtain the following expression

$$\eta_m =$$
$$\frac{2}{S_0}\frac{K_1 W}{2\pi}\frac{1}{\sigma^2}e^{-\frac{R_h^2}{2\sigma^2}}\int_{[0,\ r_1]\cup[r_2,\ R]} re^{-\frac{r^2}{2\sigma^2}}\int_0^{\pi} e^{\frac{rR_h cos(\theta)}{\sigma^2}}$$
$$\ln\left(1 + \frac{K_2}{max\left(\rho, g(r) + \alpha r^{2b}|re^{i\theta} - R_h|^{-2b}\right)}\right) drd\theta$$
$$+ \frac{2}{S_0}\frac{K_1 W}{2\pi}\frac{1}{\sigma^2}e^{-\frac{R_h^2}{2\sigma^2}}\int_{r_1}^{r_2} re^{-\frac{r^2}{2\sigma^2}}\int_{cos^{-1}(h(r))}^{\pi} e^{\frac{rR_h cos(\theta)}{\sigma^2}}$$
$$\ln\left(1 + \frac{K_2}{max\left(\rho, g(r) + \alpha r^{2b}|re^{i\theta} - R_h|^{-2b}\right)}\right) drd\theta \quad (23)$$

Following the same steps to obtain (23), the mean throughput of UEs served by the SC is

$$\eta_s = \frac{2}{S_0}\frac{K_1 W}{2\pi}\frac{1}{\sigma^2}e^{-\frac{R_h^2}{2\sigma^2}}\int_{r_1}^{r_2} re^{-\frac{r^2}{2\sigma^2}}\int_0^{cos^{-1}(h(r))} e^{\frac{rR_h cos(\theta)}{\sigma^2}} \times$$
$$\ln\left(1 + \frac{K_2}{max\left(\rho, (g(r)+1)\frac{1}{\alpha}r^{-2b}|re^{i\theta} - R_h|^{2b}\right)}\right) drd\theta \quad (24)$$

### C. Scenario 3: with introducing the impact of imperfect hotspot localization

In Scenario 3, the SC is deployed near the HS but it does not cover exactly the traffic in this HS. And so, more UEs will be served by the MC but the interference in this case will be more significant. The expression of $\eta_m$ in (23) can be easily transformed to

$$\eta_m = \frac{1}{S_0}\frac{K_1 W}{2\pi}\frac{1}{\sigma^2}e^{-\frac{R_h^2}{2\sigma^2}}\int_{[0,\ r_1]\cup[r_2,\ R]} re^{-\frac{r^2}{2\sigma^2}} \times$$
$$\int_0^{\pi}\left(e^{\frac{rR_h cos(\theta-\theta_h+\theta_s)}{\sigma^2}} + e^{\frac{rR_h cos(\theta+\theta_h-\theta_s)}{\sigma^2}}\right) \times$$
$$\ln\left(1 + \frac{K_2}{max\left(\rho, g(r) + \alpha r^{2b}|re^{i\theta} - R_s|^{-2b}\right)}\right) drd\theta$$
$$+ \frac{1}{S_0}\frac{K_1 W}{2\pi}\frac{1}{\sigma^2}e^{-\frac{R_h^2}{2\sigma^2}}\int_{r_1}^{r_2} re^{-\frac{r^2}{2\sigma^2}} \times$$
$$\int_{cos^{-1}(h(r))}^{\pi}\left(e^{\frac{rR_h cos(\theta-\theta_h+\theta_s)}{\sigma^2}} + e^{\frac{rR_h cos(\theta+\theta_h-\theta_s)}{\sigma^2}}\right) \times$$
$$\ln\left(1 + \frac{K_2}{max\left(\rho, g(r) + \frac{P_s}{P}r^{2b}|re^{i\theta} - R_s|^{-2b}\right)}\right) drd\theta \quad (25)$$

Likewise, $\eta_s$ becomes

$$\eta_s = \frac{1}{S_0}\frac{K_1 W}{2\pi}\frac{1}{\sigma^2}e^{-\frac{R_h^2}{2\sigma^2}}\int_{r_1}^{r_2} re^{-\frac{r^2}{2\sigma^2}} \times$$
$$\int_0^{cos^{-1}(h(r))}\left(e^{\frac{rR_h cos(\theta-\theta_h+\theta_s)}{\sigma^2}} + e^{\frac{rR_h cos(\theta+\theta_h-\theta_s)}{\sigma^2}}\right) \times$$
$$\ln\left(1 + \frac{K_2}{max\left(\rho, (g(r)+1)\frac{1}{\alpha}r^{-2b}|re^{i\theta} - R_s|^{2b}\right)}\right) drd\theta \quad (26)$$

## IV. NUMERICAL RESULTS

In order to assess the numerical results of the studied scenarios, we propose to realize two kind of simulations where the most important parameters are shown in Table 1.

In the first simulation, the position of the HS changes with varying $R_h$ at first and then $\theta_h$. Furthermore, we consider the error of SC positioning to be constant relative to the position of the HS. We fix an error of 0 meters (perfect HS localization), 60 meters (accuracy provided in [12], [13]) and 120 meters (current accuracy when using probes) respectively between the variables $R_h$ and $R_s$ and we suppose that $\theta_s$ is equal to $\theta_h$ (equal to $\frac{\pi}{3}$). Then, the mean user throughputs and the absorption coefficients are calculated as a function of $R_h$. Next, we take $R_s = R_h = 0.4$ Km, and we consider the error between $\theta_h$ and $\theta_s$ to be equal to $0$, $\pi/6$ and $\pi/3$ respectively and we evaluate the same performance metrics as in the first part of the simulations.

In the second simulation, we fix the position of the HS

in $0.35e^{i\frac{\pi}{6}}$ and $0.52e^{i\frac{\pi}{2}}$ respectively. Then the mean user throughputs and the absorption coefficients are plotted as a function of the SC position with varying $R_s$ (with $\theta_s = \theta_h$) at first and then $\theta_s$ (with $R_s = R_h$).

Table 1: Simulation parameters.

| Macro deployment | infinite hexagonal with $\delta = 1$ Km |
|---|---|
| Association | UE associated to highest RSRP |
| Pathloss model MtoUE | $151 + 37.6 log_{10}(d_{Km})$ |
| Pathloss model StoUE | $148 + 36.7 log_{10}(d_{Km})$ |
| BS power | Macro:46dBm, Small:30dBm |
| Antenna gain with cable loss | Macro:18dBi, Small:6dBi |
| Frequency/Bandwidth | 2.6 Ghz / 20 Mhz |
| Thermal noise per Hertz | -174dBm/Hz |
| Noise figure | 8dB |
| UE category/Throughput | 3 / $\eta_0 = 98 Mbps$, $K_1 = 0.85$, $K_2 = 1.9$ |
| UE antenna gain/Body loss | 0dB / 2dB |

From Fig. 3a, we observe that deploying a SC near the MC does not generate additional capacity gains since the interference in this case is very high comparing to the SNR received either from the serving SC or MC. In fact, for a HS in position of less than 300 meters far from the MC, the evaluation of the impact of bad localization of the traffic HS is worthless and not justified because the offloading gain[3] is negative even with a perfect positioning. Hence, the deployment of a SC near the MC does not help to offload the traffic and it deteriorates the throughput in the MC. However, the deployment of a SC improves significantly the overall performance of the MC in the presence of the HS in cell edge. In the latter case, the SC still generates positive offloading gains even when its position does not match exactly with the position of the HS.

Moreover, from Fig. 3b, it is clear that when the HS is in the center of the cell, the RSRP received from the MC is often higher than the RSRP received from the SC. As a result, the absorption coefficient relative to the presence of the SC is very small even with a perfect positioning of the SC. This coefficient is more important when the HS is in the cell edge. Moreover, deploying a SC with errors in the positioning remains a useful solution to offload an important percentage of traffic located in the cell edge. We also notice that for a HS in the cell center, a small percentage of mobile locations can be offloaded by the SC but the mean throughput in the SC (denoted by $\eta_{Sc2}$) in Scenario 2 is not improved comparing to the mean throughput in Scenario 1 (denoted by $\eta_{Sc1}$ which means that the mean user throughput is calculated only in the same region as the served area by the SC in Scenario 2). This is due to the high interference in the center between the SC and the MC. This observation is deduced from the extra curves represented with triangles for $\eta_{Sc1}$ and dotted line for $\eta_{Sc2}$ in Fig. 3a).

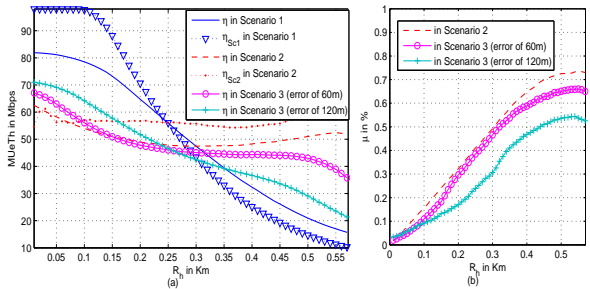

Fig. 3: The MUeTh (a) and the absorption coefficient (b) for different locations of the HS with varying $R_h$.

[3] The offloading gain is the extra capacity effectively exploited in the deployed SC and it is defined by $\rho = \frac{\eta_{Scenario\ 2,3} - \eta_{Scenario\ 1}}{\eta_{Scenario\ 1}}$

We observe in Fig. 4a that the mean throughput in Scenario 2 is higher than that in Scenario 1. However, when we introduce an error of $\pi/6$, the mean throughput becomes lower because most of the traffic is served by the MC and the SC will play the role of interfering cell. So far, when the error is taken higher than $\pi/6$, the mean user throughput is improved since the interference is more attenuated.

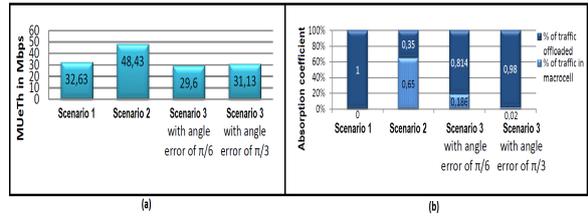

Fig. 4: The MUeTh (a) and the absorption coefficient (b) for different locations of the HS with varying $\theta_h$.

Fig. 4b shows that when $\theta_h$ is different from $\theta_s$, the absorption coefficient is reduced as compared to the case of a perfectly deployed SC, especially if this difference is high.

In Fig. 5a, we compare the mean user throughput in the presence of a traffic HS and with varying the position of the SC. When the SC is deployed near the MC, the traffic located in the cell center (HS located at $0.35Km$ far from the MC) and served by the MC will be significantly interfered by the SC. If this SC is deployed near the HS, most of the traffic will be offloaded by the SC and the mean user throughput is improved. However, we observe also that when the distance between $R_s$ and $R_h$ increases, the mean user throughput is approximately constant (function of $R_h$) and its value is equal to the mean user throughput of a network without SCs (offloading gain near 0 %). This means that when the SC is deployed far from the HS and the MC, the performance of Scenario 1 is approximately the same as that of Scenario 3.

On the other hand, we observe that when the HS is in the cell edge, the SC is an appropriate solution to offload the congesting traffic and errors of SC positioning are more tolerated comparing to the case of a HS in the center. In fact, for a traffic HS in 520m far from the MC, the SC still improve the mean user throughput in the MC even with an error of HS localization of 160m which is not the case for a HS in 350m far from the MC. In such scenarios, some traffic HS localization techniques [12], [13] can be implemented and used in operational SC planning tools.

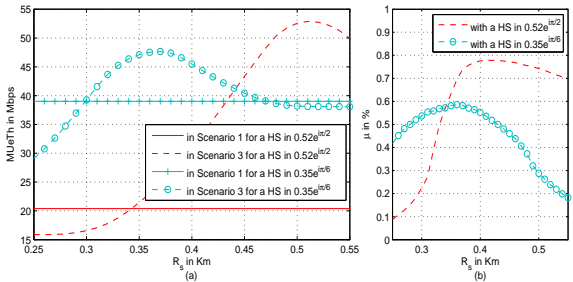

Fig. 5: The MUeTh (a) and the absorption coefficient (b) for different locations of the SC relative to the HS position with varying $R_s$.

Results in Fig. 5b show that more traffic locations are served by the SC when its coordinates approaches those of the HS. The absorption coefficient is higher for a HS in the cell edge and is reduced when the HS gets nearer to the MC.

Fig. 6a shows that when the HS is in the cell edge, the performance of the MC is improved with deploying a SC with $\theta_s - \theta_h$ less than a certain threshold. This can be explained by the fact that UEs in bad radio conditions and taken in charge by the MC are offloaded to the SC which is near to the HS. Moreover, we observe that the impact of error related to $\theta_s$ is very important and may cause significant degradation of the system performance when the distance between the HS and the MC increases. This can be explained by the fact that the distance between the SC and the HS is proportionally increased with the difference between $\theta_s$ and $\theta_h$.

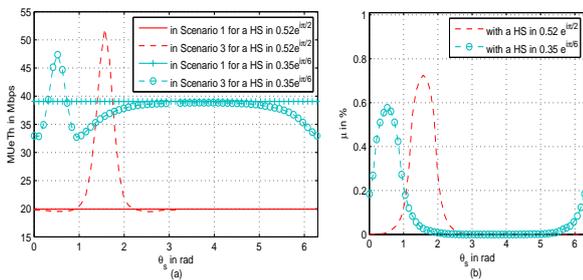

Fig. 6: The MUeTh (a) and the absorption coefficient (b) for different locations of the SC relative to the HS position with varying $\theta_s$.

Fig. 6b consolidates the different observations in Fig. 6a and it further shows that the absorption coefficient is more important when the HS is in the cell edge and the error of SC positioning is more tolerated.

## V. CONCLUSION

We studied in this paper the impact of deploying a SC in the presence of traffic HS inside a MC. Our results show that the efficiency of deploying SCs to offload traffic in the congested MC depends mainly on the HS's position within the cell as well as the SC's position with respect to the MC location. When the HS is in the cell center, even a perfect positioning of the SC is not beneficial for the user throughput and does not bring offloading gain. However, when the HS is in cell edge, errors of HS localization are more tolerated and the system performance is improved by deploying SCs as compared to a network composed of MCs only. Furthermore, our results show that for SC deployments with positive gains, the mean user throughput and the absorption coefficient essentially depend on the distance between the SC and the HS.

As a future step, we are considering analytical modeling of the system performance at the flow level, in a dynamic configuration where users arrive to the network at random time epochs and leave it after a finite service duration. Moreover, we would like to incorporate other parameters, such as shadowing, tri-sectorial sites and antenna masks.